\documentstyle[twoside,fleqn,espcrc2,epsf]{article}

\title{
Scalar Quarkonium Masses}

\author{W.Lee and D. Weingarten\\
IBM Research, P.O.~Box 218,
Yorktown Heights, NY 10598\\}

\begin{document}

\begin{abstract}

We evaluate the valence approximation to the mass of scalar quarkonium
for a range of different parameters.  Our results strongly suggest that
the infinite volume continuum limit of the mass of $s\overline{s}$
scalar quarkonium lies well below the mass of $f_J(1710)$. The resonance
$f_0(1500)$ appears to the best candidate for $s\overline{s}$ scalar
quarkonium.
 
\end{abstract}

\maketitle

For the valence approximation to the infinite volume continuum limit of
the lightest scalar glueball mass, a calculation on
GF11~\cite{Vaccarino} gives $1740 \pm 71$ MeV, favoring $f_J(1710)$ as
the lightest scalar glueball.  Among other observed resonances which
could have scalar glueball quantum numbers, only $f_0(1500)$ is near
enough to $1740 \pm 71$ MeV to be a possible alternate glueball
candidate.  Ref.~\cite{Sexton95} suggests that $f_0(1500)$ is dominantly
$s\overline{s}$ scalar quarkonium. Evidence for this identification is
given in Ref.~\cite{Weingarten96}.  As a test of the interpretation of
$f_0(1500)$ as $s\overline{s}$ scalar quarkonium, we have now
calculated, for Wilson quarks in the valence approximation, the mass of
scalar quarkonium with two different values of lattice spacing and a
range of different quark masses. Although the data we have is not
sufficient to permit an extrapolation of scalar quarkonium masses to the
continuum limit or to infinite volume, it is sufficient to show that the
continuum infinite volume value for the valence approximation to the
$s\overline{s}$ mass lies well below the mass of $f_J(1710)$.  This
result, combined with the discussion of Ref.~\cite{Weingarten96}, favors
$f_0(1500)$ as $s\overline{s}$ scalar quarkonium and makes this
interpretation for $f_J(1710)$ appear improbable.

The calculations reported here were done on the GF11 parallel computer
and required about 5 months of operation at a sustained speed of
approximation 6 Gflops.

To evaluate the mass of scalar quark-antiquark states, we define
operators for these states by first fixing each gauge configuration to
lattice Coulomb gauge, then convoluting, in the space direction, the
Wilson quark field $\Psi_{sc}(x)$ with a gaussian $G^r(\vec{x})$ with
root-mean-squared radius $r$ to produce a smeared Coulomb gauge quark
field $\Psi^r_{sc}(x)$~\cite{Butler}. Here $1 \le s \le 4$ and
$1 \le c \le 3$ are spin and color indices, respectively. 
We assume only a single flavor of quark. In a
gamma-matrix representation with $\gamma_5$ given by the
diagonal entries 1, 1, -1, -1,
define the upper and lower quark fields
\begin{eqnarray}
\label{defud}
\Psi^{ru}_c( x) & = & \sum_{s = 1, 2} \Psi^r_{sc}(x) \xi_s, \nonumber \\ 
\Psi^{rd}_c( x) & = & \sum_{s = 3, 4} \Psi^r_{sc}(x) \xi_s, \nonumber   
\end{eqnarray}
and define $\overline{\Psi}^{ru}_c(x)$ and $\overline{\Psi}^{rd}_c(x)$,
similarly from $\overline{\Psi}^r_{sc}(x)$ with $\xi_s^*$ in place of
$\xi_s$. Here the $\xi_s$, for each $s$, are independent random cube
roots of 1.  The scalar operator
\begin{eqnarray}
\label{defsxpx}
S_{\xi}^r(x) & = & \overline{\Psi}^{ru}(x) \Psi^{ru}(x)
+ \overline{\Psi}^{rd}(x) \Psi^{rd}(x),  \nonumber 
\end{eqnarray}
when averaged over $\xi$ becomes the more familiar
\begin{eqnarray}
\label{defsp}
S^r(x) & = & \overline{\Psi}^{r}(x) \Psi^{r}(x). \nonumber 
\end{eqnarray}
A pseudoscalar $P_{\xi}^r(x)$ and its average $P^r(x)$ can be
defined in analogy to $S_{\xi}^r(x)$ and $S^r(x)$.

\begin{figure}
\epsfxsize=60mm
\epsfbox{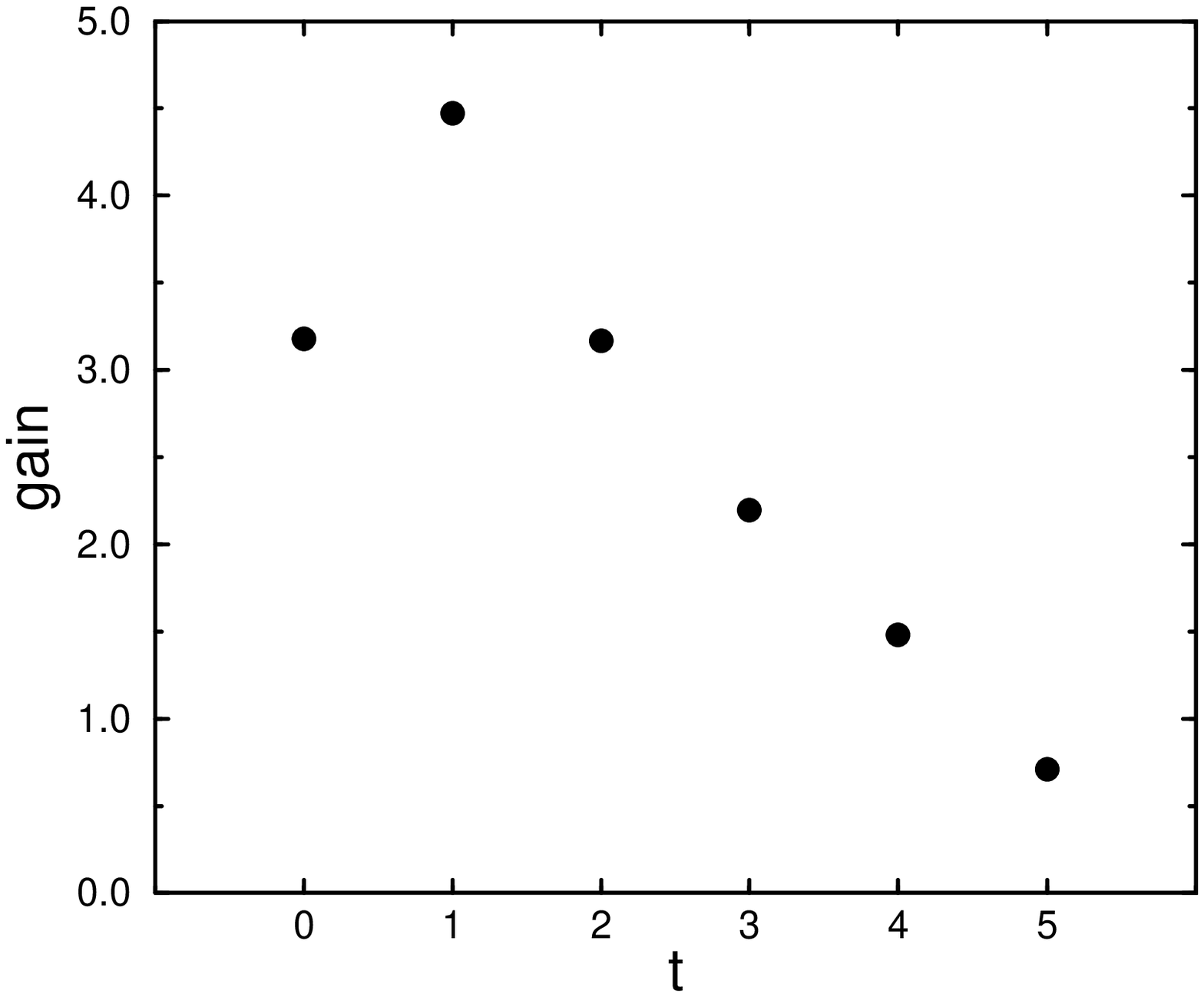}    
\vskip -12mm
\caption{
For the scalar propagator on a $16^3 \times 24$ lattice with
$\beta$ of 5.70 and $\kappa$ of 0.1650, the performance gain
of a random source in comparison to a causal source.}
\vskip -4mm
\label{fig:scgain}
\end{figure}

\begin{figure}
\epsfxsize=60mm
\epsfbox{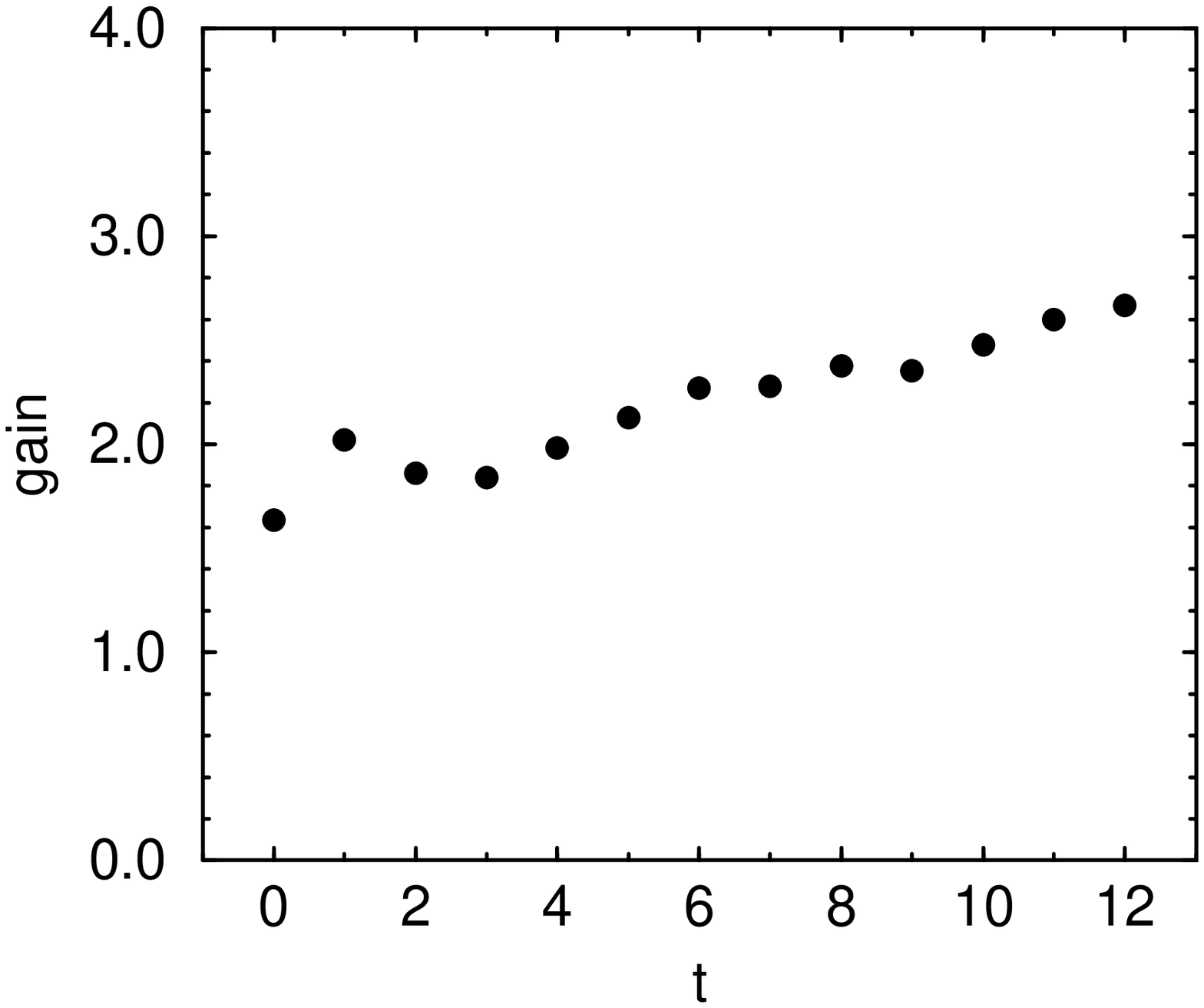}    
\vskip -12mm
\caption{
For the pseudoscalar propagator on a $16^3 \times 24$ lattice with
$\beta$ of 5.70 and $\kappa$ of 0.1650, the performance gain
of a random source in comparison to a causal source.}
\vskip -4mm
\label{fig:psgain}
\end{figure}

For scalar correlation functions we choose 
\begin{eqnarray}
C^r_S(t) & = & \sum_{\vec{x}} < S^r(\vec{x},t)
S_{\xi}^r(0) > - \nonumber \\
& & \sum_{\vec{x}} < S^r(\vec{x},t)>
<S_{\xi}^r(0) > \nonumber 
\end{eqnarray}
where averages are over $\xi$ and gauge field configurations, with one
random vector $(\xi_1, \ldots \xi_4)$ for each field. A pseudoscalar
$C^r_P(t)$ can be defined similarly.  The correlations $C^r_S(t)$ and
$C^r_P(t)$ together require a factor of six fewer quark matrix
inversions per gauge configuration than needed for $C^{\prime r}_S(t)$
and $C^{\prime r}_P(t)$ defined with $S^r(x)$ and $P^r(x)$,
respectively, used as both source and sink operators.  To obtain
propagators with a fixed statistical uncertainty, however, some of this
factor of six will be lost to the additional noise arising from $\xi$.
Figure~\ref{fig:scgain} shows the actual gain in arithmetic work for the
scalar correlation, $6 [D^{\prime}_S(t)/D_S(t)]^2$, where $D_S(t)$ and
$D^{\prime}_S(t)$ are the statistical dispersions in $C^r_S(t)$ and
$C^{\prime r}_S(t)$ respectively.  Figure~\ref{fig:psgain} shows the
corresponding actual gain in arithmetic work for the pseudoscalar
correlation.  The values in Figures~\ref{fig:scgain} and
\ref{fig:psgain} were found from 188 independent gauge configurations on
a lattice $16^3 \times 24$ with $\beta$ of 5.70 and $\kappa$ of 0.1650.
For smaller $\kappa$ we expect the gain to be greater.  For the time
intervals of our scalar mass fits, the actual gain turns out to be about
a factor of 2. Without this factor, our five months of calculation would
have required ten.

The cost of generating gauge field configurations and gauge
fixing we found could be reduced, for the scalar correlation
function, by evaluating propagators for several different source time
values on each lattice.

\begin{figure}
\epsfxsize=60mm
\epsfbox{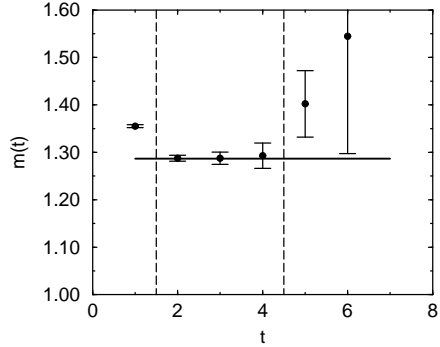}
\vskip -12mm
\caption{
Effective scalar mass, fitted mass and fitting range
for 
a $16^3 \times 24$ lattice with $\beta$ of 5.70 and
$\kappa$ of 0.1650.}
\vskip -6mm
\label{fig:efm570}
\end{figure}

\begin{figure}
\epsfxsize=60mm
\epsfbox{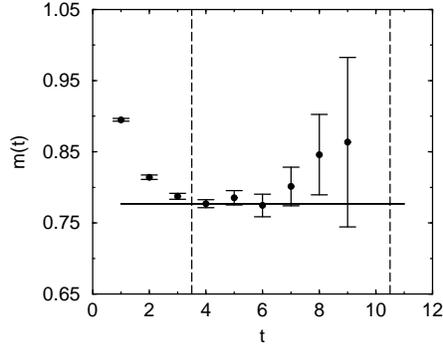}
\vskip -12mm
\caption{
Effective scalar mass, fitted mass and fitting range for 
a $24^4$ lattice with $\beta$ of 5.93 and
$\kappa$ of 0.1567.}
\vskip -4mm
\label{fig:efm593}
\end{figure}

Figure~\ref{fig:efm570} shows effective masses, the fitted mass, and the
fitting range for the scalar correlation function given by 1972 gauge
configurations on a lattice $16^3 \times 24$ with $\beta$ of 5.70,
$\kappa$ of 0.1650 and smearing radius $r$ of $\sqrt{6}$.  For each
gauge configuration, propagators were found from six different starting
times. Figures~\ref{fig:efm593} shows effective masses, the fitted mass
and the fitting range for the scalar correlation function given by 1733
gauge configurations on a lattice $24^4$ with $\beta$ of 5.93, $\kappa$
of 0.1567 and $r$ of $\sqrt{27/2}$.  For each gauge configuration,
propagators were found from four different starting times.  In physical
units, the quark mass and smearing radius in Figures~\ref{fig:efm570}
and ~\ref{fig:efm593} are nearly equal.  Scalar and pseudoscalar masses
for several different $\kappa$ at each $\beta$ are given in
Tables~\ref{tab:570} and \ref{tab:593}.  In Ref.~\cite{Butler}, the rho
mass in lattice units for $16^3 \times 24$ with $\beta$ of 5.70 and for
$24^4$ with $\beta$ of 5.93 was reported to be 0.5676(79) and
0.3851(79), respectively.  Thus in physical units the lattice period at
the two different values of $\beta$ we consider are nearly equal.

\begin{figure}
\epsfxsize=60mm
\epsfbox{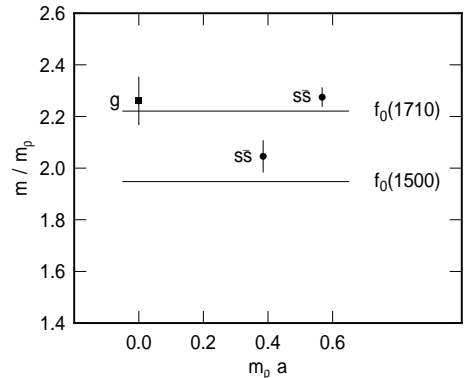}
\vskip -10mm
\caption{Scalar $s\overline{s}$ quarkonium masses for two different
values of lattice spacing.}
\vskip -4mm
\label{fig:bothbetas}
\end{figure}

The $s\overline{s}$ scalar mass we found by interpolation in quark mass
to the strange quark mass corresponding to hopping constants determined
in Ref.~\cite{Butler} to be 0.16404 at $\beta$ of 5.70 and 0.15620 at
$\beta$ of 5.93. The resulting masses, in units of
$m_{\rho}$~\cite{Butler}, are shown in Figure~\ref{fig:bothbetas} as a
function of lattice spacing in comparison to the predicted scalar
glueball mass and the masses of $f_J(1710)$ and
$f_0(1500)$. Figure~\ref{fig:bothbetas} stongly suggests that for the
fixed volume we are using, the continuum value of the $s\overline{s}$
scalar mass will lie well below the mass of $f_J(1710)$.  Since the
$s\overline{s}$ scalar has one unit of orbital angular momentum, its
physics radius should be larger than that of the $s\overline{s}$ $\phi$
meson with no orbital angular momentum.  The scalar mass divided by
$m_{\phi}$ should therefore fall with volume.  Volume dependence data
for $m_{\phi}$ in Ref.~\cite{Butler} then implies that infinite volume
values of the scalar mass divided by $m_{\rho}$ will be at most 2.8\%
above the numbers shown in Figure~\ref{fig:bothbetas}. Our qualitative
conclusion about the scalar $s\overline{s}$ mass remains the same in
infinite volume.  Thus $f_0(1500)$ appears to be a likely candidate for
the the $s\overline{s}$ scalar, while this assignment for $f_J(1710)$
appears quite unlikely.

\begin{table}
\begin{center}
\begin{tabular}{llcc} 
\hline
$\kappa$ & $m_{sc}$ & range & $\chi^2$ \\
\hline
0.1625	& 1.299(12) & 3 - 5 & 0.320 \\
0.16404 & 1.291(10) &       &       \\
0.1650	& 1.287(6) & 2 - 4 & 0.002 \\
\hline
$\kappa$ & $m_{ps}$ & range & $\chi^2$ \\
\hline
0.1625 & 0.5795(6) & 7 - 10 & 0.371 \\
0.1650 & 0.4560(5) & 7 - 10 & 0.250 \\
\hline
\end{tabular}
\caption{ 
Masses, fitting ranges and $\chi^2$ per degree of freedom of mass fits
on a $16^3 \times 24$ lattice at $\beta$ of 5.70.}
\vskip -14mm
\label{tab:570}
\end{center}
\end{table}

\begin{table}
\begin{center}
\begin{tabular}{llcc} 
\hline
$\kappa$ & $m_{sc}$ & range & $\chi^2$ \\
\hline
0.1539 & 0.856(3) & 4 - 11 & 1.39 \\
0.1554 & 0.806(4) & 4 - 11 & 1.40 \\
0.1562 & 0.788(17) &  & \\
0.1567 & 0.777(5) & 4 - 10 & 1.44 \\
\hline
$\kappa$ & $m_{ps}$ & range & $\chi^2$ \\
\hline
0.1539 & 0.4819(5) & 8 - 12 & 0.62 \\
0.1554 & 0.3982(5) & 8 - 11 & 0.26 \\
0.1567 & 0.3147(6) & 8 - 11 & 0.52 \\
\hline
\end{tabular}
\caption{ Masses, fitting
ranges and $\chi^2$ per degree of freedom of mass fits on
on a $24^4$ lattice at $\beta$ of 5.93.}
\vskip -14mm
\label{tab:593}
\end{center}
\end{table}

\end{document}